\documentclass[twocolumn,showpacs,preprintnumbers,amsmath,amssymb]{revtex4}
\usepackage{tabularx}
\usepackage{amsmath,amssymb}
\usepackage{graphicx}
\usepackage{dcolumn}
\def\apj #1 #2 #3 {#1, ApJ, {\bf #2}, #3}
\def\apjl #1 #2 #3 {#1, ApJ, {\bf #2}, L#3}
\def\apjs #1 #2 #3 {#1, ApJS, {\bf #2}, #3}
\def\aap  #1 #2 #3 {#1, A\&A, {\bf #2}, #3}
\def\mnras #1 #2 #3 {#1, MNRAS, {\bf #2}, #3}
\def\pra #1 #2 #3 {#1, Phys.~Rev.~A., {\bf #2}, #3}
\def\prb #1 #2 #3 {#1, Phys.~Rev.~B., {\bf #2}, #3}
\def\prc #1 #2 #3 {#1, Phys.~Rev.~C., {\bf #2}, #3}
\def\prd #1 #2 #3 {#1, Phys.~Rev.~D., {\bf #2}, #3}
\def\pre #1 #2 #3 {#1, Phys.~Rev.~E., {\bf #2}, #3}
\def\prl #1 #2 #3 {#1, Phys.~Rev.~Lett., {\bf #2}, #3}
\def\plb #1 #2 #3 {#1, Phys.~Lett.~B., {\bf #2}, #3}
\def\science #1 #2 #3 {#1, Science., {\bf #2}, #3}
\def\nature #1 #2 #3 {#1, Nature., {\bf #2}, #3}
\def\nphysa #1 #2 #3 {#1, Nucl.~Phys.~A., {\bf #2}, #3}
\def\nphysb #1 #2 #3 {#1, Nucl.~Phys.~B., {\bf #2}, #3}
\def\nphysbs #1 #2 #3 {#1, Nucl.~Phys.~B.~Suppl., {\bf #2}, #3}

\def\h#1{\hbox{${}^{#1}$H}}

\def\h502{\hbox{$ h^{2}_{50}$}}

\def\fun#1#2{\lower3.6pt\vbox{\baselineskip0pt\lineskip.9pt
  \ialign{$\mathsurround=0pt#1\hfil##\hfil$\crcr#2\crcr\sim\crcr}}}
\begin{document}
%
\title{Bulk Viscosity, Decaying Dark Matter, and the Cosmic Acceleration}
\author{James R. Wilson}
\affiliation{
Lawrence Livermore National Laboratory, Livermore, CA 94550}
\author{Grant J. Mathews}
\affiliation{
Center for Astrophysics, Department of Physics, University of Notre Dame, Notre Dame, IN 46556 }
\author{George M. Fuller}
\affiliation{Department of Physics, University of California, San Diego, La Jolla, CA 92092-0319}%
\date{\today}
\begin{abstract}
 We discuss a cosmology in which   cold dark-matter  particles decay into relativistic particles.  We argue that such decays  could lead naturally to a  bulk viscosity  in the cosmic fluid.   For decay lifetimes comparable to the present hubble age, this bulk viscosity
  enters the cosmic energy equation as an effective negative pressure.  We investigate whether this negative pressure is  of sufficient magnitude to
  account fo the observed cosmic acceleration.  We show that a  single decaying species in a $\Lambda = 0$, flat,  dark-matter dominated cosmology can not reproduce the observed magnitude-redshift relation from Type Ia supernovae.  However, a delayed bulk viscosity, possibly due to a cascade of decaying particles may be able to account for a significant fraction of the apparent cosmic acceleration.
  Possible candidate nonrelativistic particles for this scenario include  sterile neutrinos or 
  gauge-mediated decaying supersymmetric particles.  
   \end{abstract}
\pacs{ 98.80.Cq, 98.80.-k, 95.30.Cq, 95.35.+d, 95.36.+x, 98.80.Es, 98.62.Py}
\maketitle
%
%
%

\section{INTRODUCTION}

A significant  challenge facing modern cosmology is that of 
understanding the nature and  origin 
of both the dark energy responsible for the present apparent acceleration \cite{garnavich} and the  dark matter \cite{Feng06} responsible for most of the gravitational mass of galaxies and clusters.  
The simplest particle physics explanations for the dark matter is, perhaps, that of a weakly interacting massive particle such as the lightest supersymmetric particle, an axion, or an electroweak singlet  (e.g. "sterile" neutrino).   The dark energy, on the other hand is generally attributed to a cosmological constant, a vacuum energy in the form of a "quintessence" scalar field
possibly very slowly evolving along an effective potential, or even relativistic effects derived from the  deviation of the present matter distribution from Friedmann homogeneity \cite{kolb05}.  See \cite{Bean05} for  recent review.

  In addition to these explanations, however, the simple coincidence that  both of these unknown entities currently contribute  comparable mass energy toward the closure of the universe begs the question as to whether they could be different manifestations of the same physical phenomenon.  Indeed, many suggestions along this line have been made for so-called unified dark-matter.  One possibility is a dark matter composed of a generalized Chaplygin gas \cite{Chaplygin} for which pressure depends upon density  $p = -A/\rho^\alpha$.  Although, it has been shown \cite{Sandvik02} that a generalized Chaplygin gas   produces an exponential blow up of the matter power spectrum which is inconsistent with observations, there are also more exotic proposals such as the flow of dark matter from a higher dimension  \cite{Umezu},  or that the quintessence field itself can act as dark matter as in the
   Born-Infeld \cite{Abramo04} model.   

The possibility of particular interest for the present work, however,  is that  
of a bulk viscosity 
 within the cosmic fluid (e.g.~\cite{Sawyer06}).  Such a term  resists the cosmic expansion and therefore acts as a negative pressure.   Indeed, it has been shown
 \cite{Fabris}  that for the right viscosity coefficient,  an accelerating cosmology can be achieved 
 without the need for a cosmological constant. 
 
  Although cosmic bulk viscosity is a viable candidate for dark-energy, to date there has been
 no suggestion of how it could originate from known physics and known particle properties.
  In this paper we consider a simple mechanism for the formation of bulk viscosity by the decay of a
   dark matter particle into relativistic products.  Such decays heat the cosmic fluid and lead to an increase in entropy and are inherently dissipative in nature.  Moreover, they lead to a cosmic fluid which is out of pressure and temperature
   equilibrium and can therefore be represented by a bulk viscosity.  We propose a form for this viscosity and show that decay lifetimes comparable to the present hubble time naturally produce 
   an accelerating cosmology in the present epoch.
   
  In the next section we summarize the general form for the bulk viscosity.  Following that we consider its affect on cosmology and suggest a form for the  bulk viscosity induced by particle decay.  In the following section we discuss constraints on the properties of such particles and argue that several candidates exist.
  In Section \ref{results}  we compute  the magnitude-redshift relation for Type Ia supernovae in this cosmology
and show that a single decay does not reproduce these data.  Only if decays are delayed, e.g. by a cascade of particle  decays, can a significant fraction of the cosmic acceleration be explained in this scenario.

\section{Bulk Viscosity from Decaying Dark Matter}

The fundamental problem that we want to address is the effect of the decay of non-relativistic dark-matter particles into relativistic neutrinos.  Bulk viscosity is a way to introduce the effects of this decay on the
equations for cosmic expansion.   The physical origin of bulk viscosity in a system can be traced to deviations from local thermodynamic equilibrium.  This can be illustrated with a simple abstract example.

Suppose that the energy-momentum tensor in an expanding volume has contributions from both a component of particles obeying non-relativistic kinematics and a component following relativistic kinematics.  Imagine that in a time step the system expands, but the momenta of the relativistic and non-relativistic particles redshift (i.e.~change) differently.  In effect, this causes these two components to have different "temperatures" describing their energy-momentum distribution functions.

  The second law of thermodynamics tells us \cite{Landau,Weinberg71} that the re-establishment of thermal equilibrium through (particle decay or) scattering  of these component particles off each other or on another medium is a dissipative process that will generate entropy.  This entropy generation can be related to the expansion rate or the local fluid velocity through a bulk viscosity term.

Thus, bulk viscosity arises any time a fluid expands to rapidly and ceases to be in thermodynamic equilibrium .  The bulk viscosity, therefore, is a measure of the pressure required to restore equilibrium to a compressed or expanding system \cite{Okumura03, Ilg99,Xinzhong01}.  Hence, it is natural for such a term to exist in the cosmologically expanding universe anytime the fluid is out of equilibrium.    Usually, in cosmology the restoration processes are taken to be so rapid that the establishment of equilibrium is almost immediate.  However,  
there is a finite time for the system to adjust to the change of the equation of state induced by particle decays.  
For the cosmology proposed here, the attainment of equilibrium as the universe expands is delayed by the gradual decay of one or more species to another which occurs over $\sim 10^{10}$ yrs.  This leads to nontrivial dependence of pressure on density as the universe expands, and therefore a bulk viscosity.

 To see how this enters quantitatively in cosmology,  we begin by summarizing  the general treatment of imperfect fluids of Weinberg \cite{Weinberg71}.  It will provide further insight into the nature of the bulk viscosity.  

When a fluid  expands (or is compressed) and departs from thermodynamic equilibrium the processes that restore equilibrium are irreversible.  Hence, they are in general accompanied by an increase in  entropy which is evidenced in the dissipation of energy.  For the case of interest here,
the increase in entropy and dissipation is the heating and pressure produced by the particle decays.
The existence of such dissipation leads to a modification of the perfect-fluid energy-momentum tenor,
\begin{equation}
T^{\mu \nu} = (\rho +  p) U^\mu U_\nu + g^{\mu \nu}  p  + \Delta T^{\mu \nu}~~,
\end{equation}
where $ \rho$ and $p$ denote density and pressure while $U^i$ is the four velocity. 
Processes of  heat flow and shear can play no role in the Friedmann-Lemaitre-Robertson-Walker (FRLW) homogeneous and isotropic conditions  of interest here.   Hence, the 
 only possible non-adiabatic  dissipative contribution $ \Delta T^{\mu \nu}$ which guarantees translational and rotational invariance for a fluid in motion with four velocity $U^\nu$ is given by \cite{Weinberg71} 
\begin{equation}
\Delta T^{\mu \nu} = -\zeta 3 \frac{\dot a}{a} \biggl(g^{\mu \nu} + U^\mu U^\nu \biggr) ~~,
\end{equation}
where $a$ is the cosmic scale factor as specified below and $\zeta$ is the bulk viscosity coefficient.  The total energy-momentum tensor is
\begin{equation}
T^{\mu \nu} = \biggl(\rho +  p - \zeta 3 \frac{\dot a}{a}\biggr) U^\mu U_\nu + g^{\mu \nu}  \biggl(p  - \zeta 3 \frac{\dot a}{a}\biggr)
~~.
\label{emom}
\end{equation}
From Eq. (\ref{emom}) it  is obvious that the effect of bulk viscosity is to replace the fluid pressure with an effective pressure given by,
\begin{equation}
p_{eff} = p - \zeta 3 \frac{\dot a}{a}~~.
\label{peff}
\end{equation}
Thus, for large $\zeta$ it is possible for the negative pressure term to dominate and an accelerating cosmology to ensue.  It is necessary, therefore, to clearly define the bulk viscosity for the system of interest.

 \section{Cosmology with Bulk Viscosity}
To examine the effect of the bulk viscosity from particle decay on the  cosmic acceleration, we analyze a flat ($k = 0$, $\Lambda = 0$) cosmology in a comoving FLRW metric,
 \begin{equation}
 g_{\mu \nu} dx^\mu dx^\nu = -dt^2 + a^2(t) \biggl[
{dr^2} + r^2 d \theta^2 + r^2 \sin^2{\theta} d\phi^2\biggr]~~,
\label{RW}
 \end{equation}
 for which $U^0 = 1$, $U^i = 0$, and $U^{\lambda}_{~; \lambda} = 3 \dot a/a$.

 We consider a fluid with total  mass-energy density $\rho$ given by,
\begin{equation}
\rho = \rho_{\rm DM} + \rho_{\rm b} + \rho_{\rm h} + \rho_\gamma + \rho_{\rm l}~~,
\label{rhotot}
\end{equation}
 where $\rho_{\rm b}$ is the baryon density,  $\rho_{\rm DM}$  is the contribution from stable dark matter,
 $\rho_{\rm h}$ is the density in unstable decaying dark matter,
 $\rho_{\rm l}$ is the produced relativistic energy density from decay while $\rho_\gamma$ is any other relativistic matter, i.e. photons and neutrinos from the big bang.   Because of decay, neither the total energy density in relativistic particles $\rho_{\rm r} = \rho_\gamma + \rho_{\rm l}$  nor the pressure $p = \rho_{\rm r}/3$ is negligible for this cosmology even at the present epoch.
 
 In the FLRW frame, the energy-momentum tensor (Eq.~\ref{emom}) then reduces to
 \begin{eqnarray}
 T_{0 0} &=& \rho \\
 T_{0 i} &=& 0 \\
 T_{i j} &=& \biggl(p - 3 \zeta \frac{\dot a}{a}\biggr) g_{i j}~~.
 \label{TFRW}
 \end{eqnarray}
 where again, this last equation shows that the bulk viscosity enters as an effective negative pressure (i.e. dark energy) in the energy-momentum tensor.
 
  The Friedmann equation does not depend upon the effective pressure and is exactly the same as for a non-dissipative cosmology, i.e.
  \begin{equation} 
H^2 = \biggl( \frac{\dot a}{a}\biggr)^2=\frac{8}{3}\pi G \rho ~~, 
\label{Friedmann}
\end{equation}
where 
$\rho$ is the  total mass-energy density from matter and relativistic particles (Eq.~\ref{rhotot}).

Although absent from the Friedmann equation, the bulk viscosity does appear in the conservation 
condition $T^{\mu \nu}_{~~;\nu} = 0$.  To illustrate this consider a flat $k = 0$, $\Lambda = 0$ cosmology
and ignore the small contribution from $\rho_{\rm b}$ and initial background radiation $\rho_\gamma$.
\begin{equation}
\rho =  \rho_{\rm h} + \rho_{\rm l} 
\end{equation}

 The conservation  equations can be solved to give the energy densities in matter and radiation:
 \begin{equation}
 \rho_{\rm h} =  \frac{1}{a^3} \rho_{m0} e^{- t/\tau}~~.
 \end{equation}
 and
 \begin{equation}
  \rho_{\rm l} = \frac{1}{a^4} \biggl[ \rho_{l0} +  \frac{ \rho_{h0}}{\tau} \int_0^t e^{- t'/\tau}a(t')dt'
  +  \rho_{BV} \biggr]~~,
  \end{equation}
  where $\rho_{BV}$ is the dissipated energy in light relativistic species due to the cosmic
  bulk viscosity,
 \begin{equation}
  \rho_{BV}= 9 \int_0^t  \zeta(t') \biggl(\frac{\dot a}{a}\biggr)^2 a(t')^4 dt'~~.
  \end{equation}
  The total density for the Friedmann equation will then include not only terms from
  heavy and light dark matter, but  a dissipated energy density in bulk viscosity.
  This is the term that contributes to  the cosmic acceleration.

\subsection{Bulk Viscosity Coefficient}

Bulk viscosity can be thought of \cite{Landau, Weinberg71, Okumura03} as a relaxation phenomenon.  It derives from the fact  that the fluid 
requires time to restore its equilibrium pressure from a departure which occurs during expansion.  
The viscosity coefficient $\zeta$ depends upon the difference between the pressure $\tilde p$ of a fluid being compressed or expanded and the pressure $p$ of  a constant volume system in equilibrium.
Of the several formulations  \cite{Okumura03} the basic non-equilibrium method \cite{Hoover80} is
identical \cite{Weinberg71} with Eq.~(\ref{peff}).
\begin{equation}
\zeta 3 \frac{\dot a}{a} = \Delta p~~,
\label{zeta1}
\end{equation}
where $\Delta p = \tilde p - p$ is the difference between the constant volume equilibrium pressure and the actual fluid pressure.  

In  Ref.~\cite{Weinberg71} the bulk viscosity coefficient is derived for a gas in thermodynamic equilibrium at a temperature $T_M$ into which radiation is injected with a temperature $T$ and a mean thermal equilibration time $\tau_{\rm e}$.  The solution for the relativistic transport equation \cite{Thomas30} can then be used to deduce infer \cite{Weinberg71} the bulk viscosity coefficient.
For this case the form  of the pressure deficit and associated bulk viscosity can be deduced from Eq.~(2.31) of Ref.~\cite{Weinberg71} which we modify slightly  and write as,
\begin{equation}
\Delta p  \sim \biggl(\frac{\partial p}{\partial T}\biggr)_n (T_M - T) = \frac{4 \rho_\gamma \tau_{\rm e}}{3}
\biggl[ 1 - \biggl(\frac{3\partial p}{\partial \rho}\biggr)\biggr] \frac{\partial U^\alpha}{\partial x^\alpha} ~~,
\label{Weinbv}
\end{equation}
where the factor of 4 comes from the derivative of the  radiation pressure $p \sim T^4$ of the  injected gas, and the term in brackets derives from the detailed solution to the linearized relativistic transport equation \cite{Thomas30}.  This term guarantees that no bulk viscosity can exist for a completely relativistic gas.
In the cosmic fluid, however, we must consider a  total  mass-energy density $\rho$ given by both nonrelativistic and relativistic components.  
 
In the present context we also have a thermalized gas into which relativistic particles at some effective temperature are being injected.  The deficit from equilibrium pressure, however,  is due to the presence of unstable decaying nonrelativistic dark matter. At any time in the cosmic expansion the pressure deficit will be  1/3 of the remaining 
mass-energy  density 
of unstable heavy particles.  Hence, we replace $\rho_\gamma/3$ with $\rho_{\rm h}/3$ in  Eq.~(\ref{Weinbv}) and write,
\begin{equation}
 \Delta p =  \frac{4{\rho_{\rm h} \tau_{\rm e}}}{3}
\biggl[ 1 - \biggl(3 \frac{\partial p}{\partial \rho}\biggr)\biggr]  \frac{\partial U^\alpha}{\partial x^\alpha}~~.
\label{Weinbv2}
\end{equation}
Here, the equilibration time $\tau_{\rm e}$ is determined  \cite{Okumura03} from the particle decay time $\tau$,
\begin{equation}
\tau_{\rm e}  = \int_0^\infty \frac{\Delta p(t)}{\Delta p(0)} dt = \frac{\tau}{[1 - 3 (\dot a/a)\tau]}
\end{equation}
where $\Delta p(0)$ denotes the initial pressure  and the denominator results from  approximating  $H = \dot a/a \approx $ constant. Note, that this factor  acts as a limiter to prevent
unrealistically large bulk viscosity in the limit of a large $\tau$.

  Following the derivation in \cite{Weinberg71}, and inserting  Eq.~(\ref{Weinbv2})  in place of Eq.~(\ref{Weinbv}), we infer the following ansatz for the bulk viscosity of the cosmic fluid due to particle decay,
\begin{equation}
\zeta = \rho_{\rm h} \tau_{\rm e} \biggl[ 1 - \frac{\rho_{\rm l} + \rho_{\gamma}}{\rho} \biggr]^2
~~,
\label{zeta}
\end{equation} 
where the square of the term in brackets comes from inserting Eq.~(\ref{Weinbv2}) into the linearized relativistic transport equation of Ref.~\cite{Thomas30}. 
   Equation (\ref{zeta}) implies a non-vanishing bulk viscosity even in the limit of long times as long as the total mass energy density is comprised of a mixture of relativistic and nonrelativistic particles.
 Hence, one should be cautious about using this linearized approximation in the long lifetime limit.
 Even so,  a more general derivation
has been made \cite{Xinzhong01} which shows that, even in the limit of interest here of a long radiation equilibration time  there is a non-vanishing bulk viscosity consistent with experimental determinations.

  \section{Decaying Dark Matter Candidates}
 
Having postulated the existence of a decaying dark matter particle, it is important to
briefly examine the constraints on such decays and whether such candidate particles could exist.
To avoid observational constraints  the decay products must have very little energy in photons or charged particles.  
The implied background in energetic photons with an energy density comparable to
the present matter energy density would have been easily detectable.  Hence, the decay products must
be in some form which is not easily detectable.  Neutrinos would be a natural candidate for
such a background.   In this case there are several decaying dark matter possibilities which come to mind.

\subsection{ Sterile Neutrinos}

Perhaps the most realistic  possibility is the decay of a  sterile neutrino into light 
 "active" neutrinos 
\cite{AFP}.  
Models have been proposed in which singlet "sterile" neutrinos 
$\nu_{\rm s}$ which mix in vacuum with active neutrinos 
($\nu_{\rm e}$, $\nu_\mu$, $\nu_\tau$) provide warm and cold dark matter 
candidates \cite{DW,SF,DH,AFP,AF,AKS, A1}. 
In most of these models the sterile neutrinos are produced in the 
very early universe through active neutrino scattering-induced de-coherence. 
This process could be augmented by medium enhancement stemming from 
a significant lepton number. In these sterile neutrino 
production processes there are two principal parameters: 
(1) the sterile neutrino mass $m_{\rm s}$; and (2) the sterile 
neutrino's vacuum mixing angle $\theta$ with one or more 
of the active neutrino flavors. The net lepton number(s) of the 
universe could be regarded as an additional parameter.
By virtue of the mixing with active neutrino species, the sterile 
neutrinos are not truly "sterile" and, as a result, can decay. For 
$m_{\rm s} < 10\,{\rm MeV}$ the dominant $\nu_{\rm s}$ decay mode is into light, 
active neutrino species. The rate for this process is \cite{AFP}
\begin{equation}
\Gamma_{\nu} \approx \left( 8.7\times{10}^{-21}\,{\rm s}^{-1} \right) 
{\left({{\sin^22\theta}\over{{10}^{-15}}} Ê\right)} {\left({{m_{\rm s}}\over  
{1\,{\rm MeV}}} \right)}^5.\label{rate3}
\end{equation}
Likewise, there is a sub-dominant $\nu_{\rm s}$ decay branch into a light 
active neutrino and a photon with rate
\begin{equation}
\Gamma_{\nu \gamma} \approx 
\left( 6.8\times{10}^{-23}\,{\rm s}^{-1} \right)
{\left({{\sin^22\theta}\over{{10}^{-15}}} \right)}
{\left ({{m_{\rm s}}\over{1\,{\rm MeV}}} \right)}^5.
\label{ratenuphoton}
\end{equation}
In this process the photon will be mono-energetic with an energy 
which is half the $\nu_{\rm s}$ rest mass. Because the primary $\nu_{\rm s}$ decay mode and the radiative branch scale Êhe same way with $m_{\rm s}$ and $\sin^22\theta$, 
there is a fixed ratio of these rates.
The best particle candidates for a decay- 
induced bulk viscosity  are those with a lifetime 
of order the Hubble time $H_0^{-1}$ and rest masses $\sim 1\,{\rm MeV}$.
Setting $\Gamma_{\nu} = H_0$ we find that the relation between the 
$ \nu_{\rm s}$ rest mass and vacuum mixing angle is
\begin{equation}
m_{\rm s} \approx 3.1\,{\rm MeV} {\left( {{h}\over{0.71}} \right)}^{1/5} 
{\left( {{{10}^{-15}}\over{\sin^22\theta}} \right)}^{1/5},
\label{massmix}
\end{equation}
where $h$ is the Hubble parameter at the current epoch in units of 
$100\,{\rm km}\,{\rm s}^{-1}\,{\rm Mpc}^{-1}$ and we have scaled our 
result to $h=0.71$, the WMAP best fit value. 
We conclude that one or 
a number of sterile neutrinos with rest masses in the $\sim {\rm MeV} $
 range could provide a significant
 decay-induced bulk viscosity. 
 
 Regarding  observational constraints,
let us note that our bulk viscosity-selected range for $m_{\rm s}$ from Eq. 
\ (\ref{massmix}) is relatively insensitive to the vacuum mixing angle. 
However, the radiative decay branch rate $\Gamma_{\nu \gamma}$ is 
linearly proportional to $\sin^22\theta$. Keeping $m_{\rm s} \sim 1\,{\rm MeV}$, 
we can adjust $\sin^22\theta$ so that the diffuse decay photon 
flux is just at or below the observational limit \cite{RT,AFP,AFT} from the 
Diffuse Extragalactic Background Radiation (DEBRA). 
For this $m_{\rm s}=1\, {\rm MeV}$ the 
DEBRA limit would require $\sin^22\theta \le {10}^{-15} $. 

We conclude that it is possible to meet the bulk viscosity 
lifetime requirement and (barely) get under the DEBRA limit with sterile neutrinos
as the decaying dark matter. 
We also note that sterile neutrinos with these parameters 
($m_{\rm s} \approx 1\,{\rm MeV}$, $\sin^22\theta \approx {10}^{-15}$) 
could be produced in the early universe in the requisite 
relic densities ({\it i.e.,} near closure) only in scenarios with 
large lepton number(s) and medium-enhanced de-coherence \cite{AFP,A2}, or with new neutrino couplings \cite{AKS}.
 
 \subsection{Decaying Supersymmetric Dark Matter}
 For supersymmetric dark matter candidates, It is generally assumed \cite{Feng06} that the initially produced dark
matter relic must be a superWIMP  in order to produce the correct relic density.
Later, this superWIMP is then presumed  to decay to a lighter stable dark matter particle.
One interpretation of such a candidate for decaying dark matter,  is then a decaying superwimp 
with a lifetime comparable to the present hubble time.  

Alternatively, the light supersymmetric
particle itself, might a candidate for decay.
If the dark matter is a light unstable supersymmetric particle, then one might imagine an R-parity violating decay.  In one scenario a particle might decay by coupling to right-handed neutrinos which then decay to normal neutrinos.  Another possibility could be gauge-mediated supersymmetry breaking 
involving the decay of  a supersymmetric sneutrino into a gravitino  plus a light neutrino.

\section{Results}
\label{results}
 
Having defined the cosmology of interest we now examine the magnitude-redshift relation for
 Type Ia supernovae (SNIa).
The apparent brightness of the Type Ia supernova standard candle with
redshift is given \cite{Carroll} by a simple relation for a flat $\Lambda = 0$ cosmology.  The luminosity distance becomes,
\begin{eqnarray}
D_L &=& \frac{c (1+z)}{ H_0  } \biggl\{  \int_0^z dz' 
\biggl[\Omega_\gamma (z') + \Omega_{\rm l} (z')
\nonumber \\
&& 
+ (\Omega_{\rm DM}(z')
 + \Omega_{\rm b}(z') + \Omega_{\rm h}(z'))  
  \biggr]^{-1/2}  \biggr\}~~,
\end{eqnarray}
where $H_0$ is the present Hubble parameter. 
The $\Omega_i$ are the energy densities normalized by the critical density at each epoch, i.e. $\Omega_i(z) = {8 \pi G \rho_i(z) / 3 H_0^2}$. $\Omega_{\rm h}$ is the closure contribution of the decaying heavy cold dark matter particle which is taken to produce light neutrinos $ \Omega_{\rm l}$ or other relativistic particles  $\Omega_\gamma$ as it decays.
Note that $\Omega_{\rm h}$, $\Omega_\gamma$ and $ \Omega_{\rm l}$ each have a nontrivial  redshift dependence due to particle decays, while
stable dark matter and baryons $\Omega_{\rm DM}(z)
 + \Omega_{\rm b}(z) $ obey the usual $(1+z)^{-3}$ dependence with redshift.  Here, and in the following discussion we will define $\Omega_M$ as the present sum of nonrelativistic matter, i.e.~ $\Omega_{\rm M} = \Omega_{\rm h}(z=0) + \Omega_{\rm DM}(z=0)
 + \Omega_{\rm b}(z=0$.)

\begin{table}[ht]
\begin{center}
\begin{tabular}{lcccc}
\hline
 $\tau$  (Gyr)            & $\Omega_{M}$ & $\Omega_{\Lambda}$ & $\chi_r^2$  \\
 \hline
$\infty$ ($\Lambda$CDM)         & 0.31    & 0.69 & 1.14              \\          
20         & 0.16    & 0. &  3.89            \\
$1\times 6$, 20 & 0.37    & 0. &  1.90          \\
$\infty$ (CDM)         & 1.    & 0. &  3.23              \\

    \\ \hline

\end{tabular}
\end{center}
\caption{Parameter sets for various fits to the SNa luminosity-redshift relation for $H_0 = 71$ km s$^{-1}$ Mpc$^{-1}$ and $\Omega_{\rm b} = 0.044$.  In the decaying (finite $\tau$) models no stable dark matter was assumed (i.e.~$\Omega_{\rm DM} = 0$).}
\label{table_params}
\end{table}

\begin{figure}[t]
\begin{center}
\includegraphics[width=8.cm]{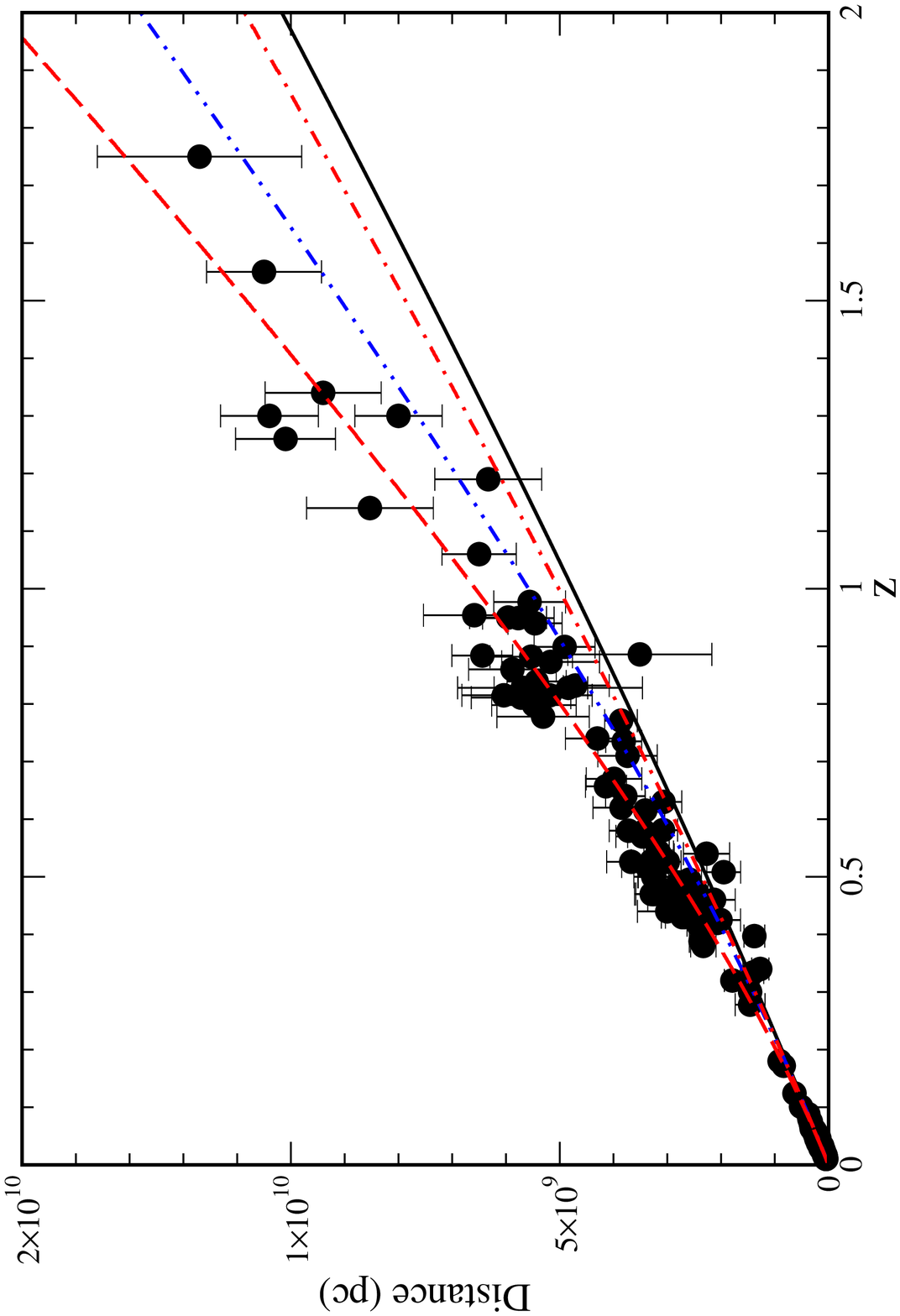}
\includegraphics[width=8.cm]{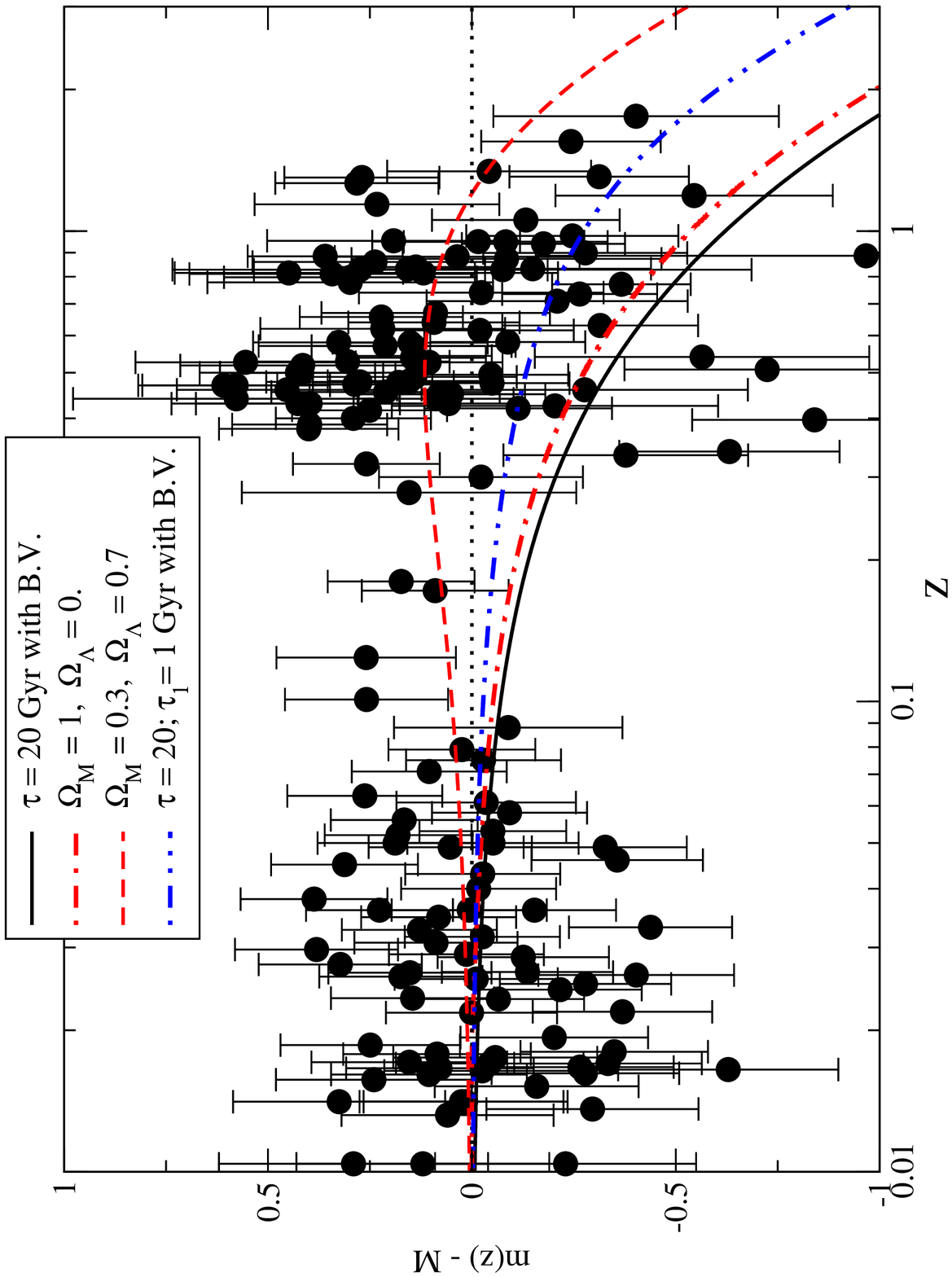}
\end{center}
\vspace{-0.5cm}
\caption{Evolution of luminosity distance with redshift for a cosmology with bulk viscosity. Points are from the Gold data set of \cite{Riess}. The upper figure shows the luminosity distance vs. redshift.  The lower figure shows the evolution of magnitudes relative to a fiducial $\Omega_k = 1$ open cosmology.
In each figure the upper dashed line shows the evolution of a standard $\Lambda$CDM cosmology and lower dot-dashed line shows the evolution of an $\Omega_M = 1$ cosmology.
The solid line is for a illustrative decaying dark matter model with $\tau = 20$ Gyr.
the dash-dot-dot line illustrates  the evolution of a cosmology in which a cascade of six particle decays each with a lifetime of $\tau_1 =1$ Gyr is followed by a final radiative decay with $\tau = 20$ Gyr.     }
\label{fig:mag}
\end{figure}

Figure \ref{fig:mag} compares various cosmological models with some  of the recent combined data from the High-Z Supernova Search Team
 and the Supernova Cosmology Project
\cite{garnavich,Riess}, while Table 1 summarizes the relevant parameters and reduced $\chi^2$ goodness of fit.
The lower figure shows
the K-corrected magnitudes  $m = M + 5 \log{ D_L} + 25$
vs.~redshift plotted relative to an open $\Omega_{rm DM},
\Omega_B , \Omega_{\Lambda}=0$,  $\Omega_k  = 1$
cosmology. 

The solid line on the upper and lower graphs in Figure \ref{fig:mag} shows the result of adding bulk viscosity from particle decay.  The upper figure gives the distance-redshift relation while the lower figure  shows the evolution of magnitudes relative to a fiducial $\Omega_k = 1/(a_0 H_0)^2 = 1$ open cosmology,
for which
\begin{eqnarray}
D_L(\Omega_k=1)&=& \frac{c (1+z)}{ 2 H_0  }  \biggl[z + 1 - \frac{1}{(z+1)}\biggr]~~,
\end{eqnarray}
and the relative distance modulus  is given in the usual way  $\Delta (m - M) = 5 \log{[D_L/D_L(\Omega_k=1)]}$.

\begin{figure}[t]
\begin{center}
\includegraphics[width=8.cm]{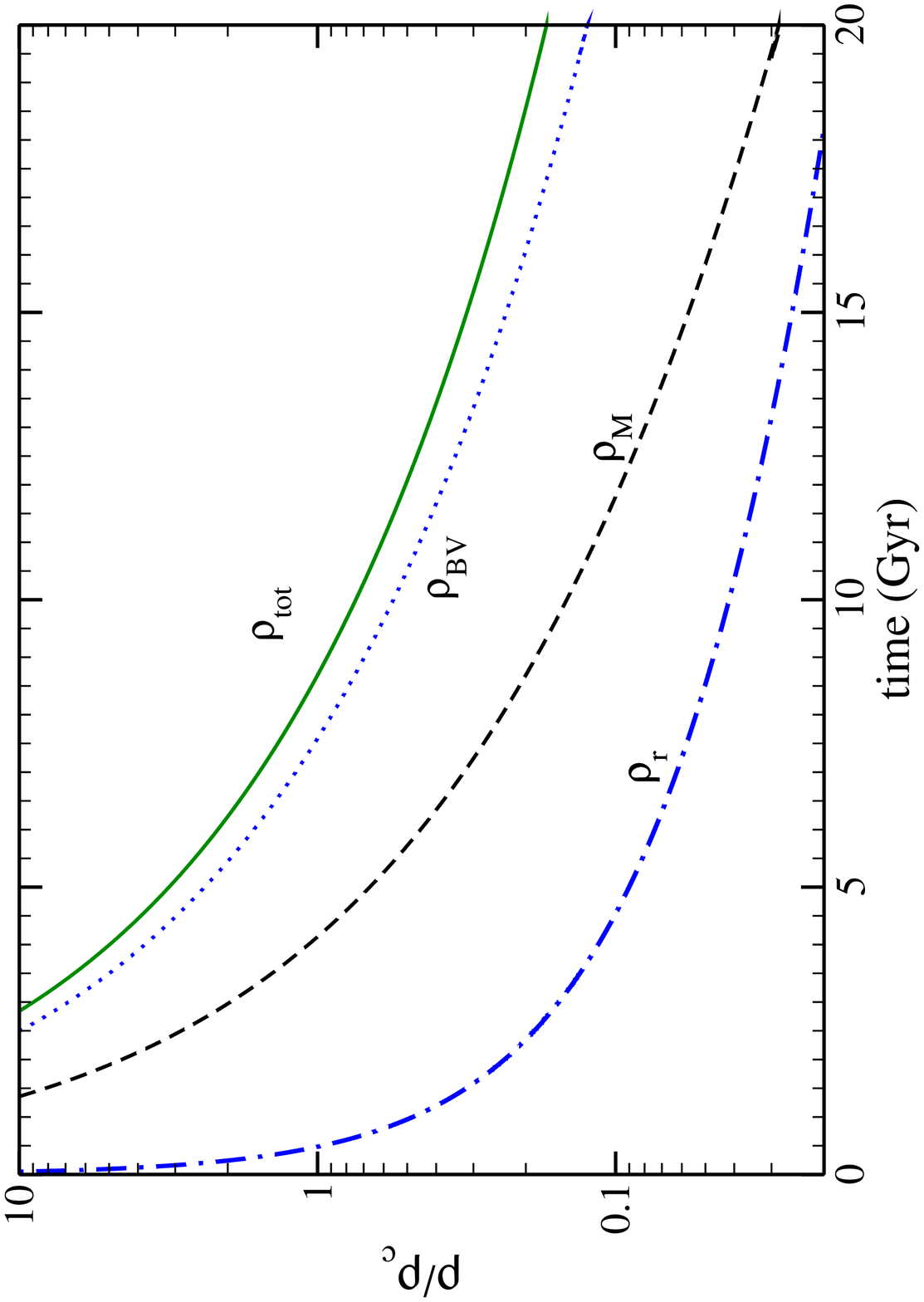}
\end{center}
\vspace{-0.5cm}
\caption{Evolution of the quantities $\rho_\gamma$, $\rho_M$, $\rho_{BV}$, and $\rho_{tot}$
as labeled for a cosmology in which the dark matter decays with a lifetime of 20 Gyr}
\label{rhobv1}
\end{figure}

From the lower graph of Figure \ref{fig:mag} we see that, although the bulk viscosity  has indeed provided  a negative pressure it does not reproduce the supernova
distance-red shift relation.  In fact it is much worse than the usual $\Lambda$CDM cosmology and is even worse than a  pure matter dominated cosmology.
The reason for this can be discerned from Figure \ref{rhobv1}.
Although the bulk viscosity is substantial, it scales with the decaying dark matter which falls 
off faster with time than $a^{-3}$ because of the decay.  An accelerating cosmology requires 
a nearly constant value of $\rho_{tot}$ with time.

A flattening of $\rho_{tot}$ could be achieved in this context if the onset of the bulk viscosity could  be
delayed until near the present epoch due to  a cascading decay.  In this possibility, the final decay would produce the relativistic products and the bulk viscosity.  This final decay, however  is preceded by a series of decays to nearly degenerate states with shorter lifetime.  

The cascade possibility might occur, for example, among sterile neutrinos.  Another cascade possibility  \cite{Feng06} is that the that the initially produced dark
matter relic is a superWIMP.
Later this superWIMP decays to a lighter stable dark matter particle, e.g.~a gravitino.
One could have a cascade of  decaying  superWIMP states  to a final unstable state or to 
unstable  light supersymmetric
particle states.

\begin{figure}[t]
\begin{center}
\includegraphics[width=8.cm]{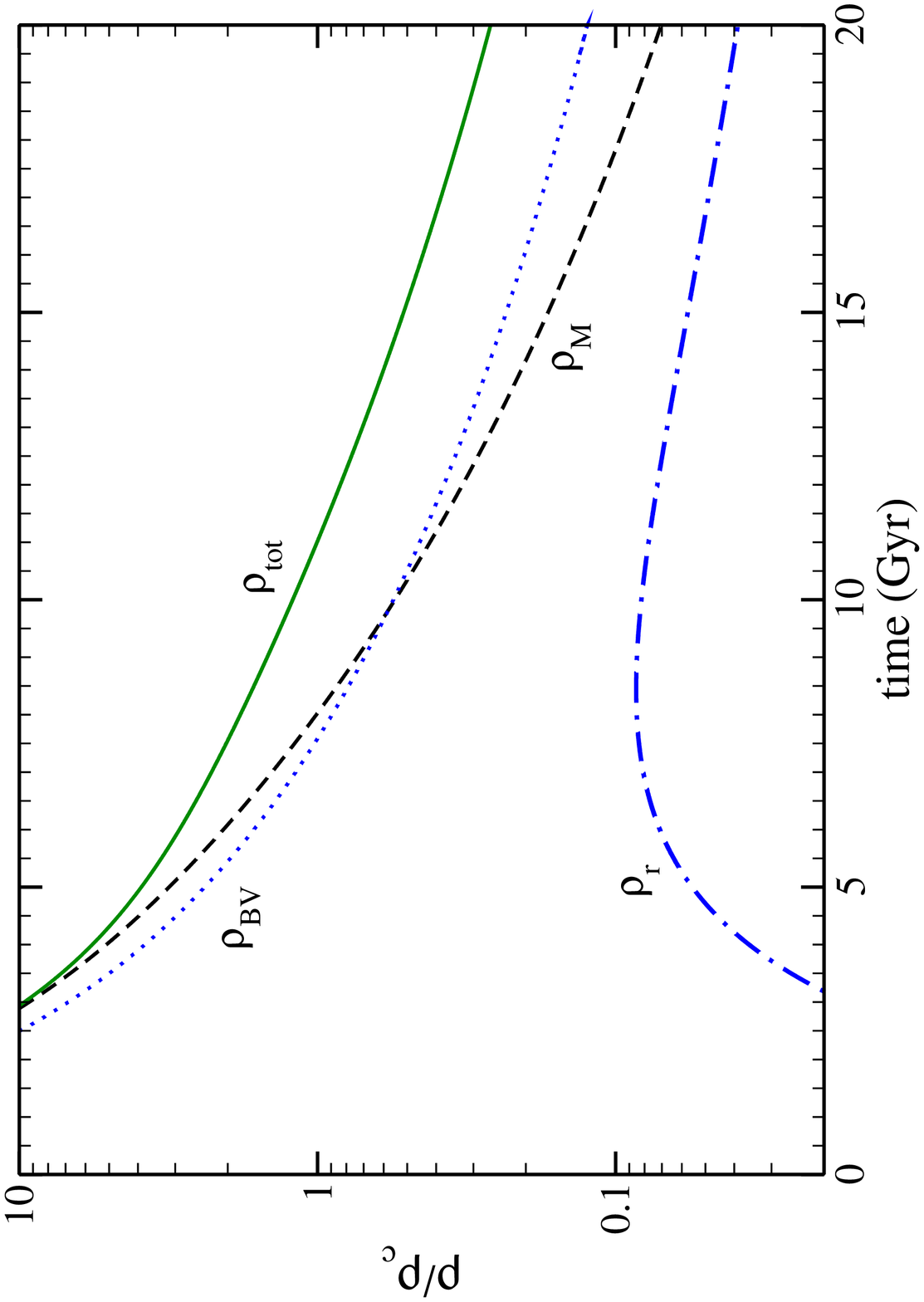}
\end{center}
\vspace{-0.5cm}
\caption{Evolution of the quantities $\rho_\gamma$, $\rho_M$, $\rho_{BV}$, and $\rho_{tot}$
as labeled in a cosmology in which decays among six nearly degenerate states occur with a lifetime of 
$\tau_1$ 1 Gyr each is followed by the decay of a long lived particle with $\tau = 20$ Gyr.}
\label{rhobvmult}
\end{figure}

Figure \ref{rhobvmult} illustrates a possible  evolution of energy density in this scenario.
Here, a decay among six states each with a lifetime of $\tau_1$ 1 Gyr each is followed by the decay of a long lived particle with $\tau = 20$ Gyr
for the case in which all matter starts in the first member of the cascade. The activity in the final decay product is delayed by the time needed to
decay through the intervening states.  The rates of the final and intermediate decays decay are given by a solution to the Bateman equation whereby the abundance  of the final product is given by
\begin{equation}
\rho_{\rm h} = \Sigma {h_j \exp{(-\lambda_j t)}}
\end{equation}
where $\lambda_j = \tau_j^{-1}$ is the decay rate of each species and the $h_j$ are given by,
\begin{eqnarray}
h_j &= & \Pi_{i \ne j}\biggl[\frac{ \lambda_j}{(\lambda_i-\lambda_j)}\biggr]~~,
\end{eqnarray}

For this possibility, we found that it was not possible in this way to completely account for the cosmic acceleration, though a significant fraction could be obtained as illustrated by the dash-dot-dot lines on
Figure \ref{fig:mag}.

\section{Conclusion}
We have considered models in which the apparent cosmic acceleration is affected by the 
bulk viscosity produced from the decay of a  dark matter particle to  light relativistic species. 
An expression for the bulk viscosity is deduced and the implied redshift-distance relation has been computed.   

As an illustrative example we considered the decay of dark matter with a lifetime of 20  Gyr
 in this cosmology.
  From the reduced $\chi^2_r$ values in Table 1, and the lines in Figure \ref{fig:mag} it is apparent that a flat $\Lambda=0$  cosmology with bulk viscosity from decay of a single dark-matter species does not
  do better than a $\Lambda$CDM or a matter dominated cosmology.  This is because the total mass-energy density does not become nearly constant with scale factor, but falls off more rapidly than even a simple matter dominated cosmology due to the combined effects of the decay of the dark matter and the emergence of a high density of
  relativistic particles.  We show, however that if the emergence of the bulk viscosity is delayed, then some, but not all, of the acceleration required by
observations of type Ia supernovae at high redshift can be explained.  As we have outlined above, one mechanism for delaying the bulk viscosity could be  a cascading decay process.

Obviously, however, one must decide whether the dilemma of a cosmological constant is less plausible than the dilemma of bulk viscosity produced by a delayed cascade of decaying dark matter particles. Our goal here, however, has merely been to argue that the possibility exists.   Having established that at least a possible paradigm exists, in future work we will examine the possible influence of this scenario on the CMB and the growth of large scale structure  which will also constrain this possibility.

Indeed, a number of recent studies (e.g. \cite{kolb05}) suggest that changes in the extrinsic curvature due to changing relativistic gravity in an inhomogeneous cosmology can lead to cosmic acceleration.
We are currently developing a computer
model similar to \cite{Centrella84} that will include the decay of
heavy neutrinos.  Our preliminary analysis indicates that  the cosmic acceleration in a universe with this type of dark matter decay may be enhanced by its effect on large scale structure, i.e.  
        decay of heavy neutrinos in a non
homogeneous cosmos may increase the expansion
effect.  

In brief, the decay produces a flow of  light neutrinos from galactic clusters.  
Given a decay time $\tau$
 and galactic clusters separated by a distance $L$
 this flow produces a  
momentum density of
 the order $s=\rho L/\tau$ (in units of $c=G=1$).
 From the momentum
constraint  [Eq.~(A4) of \cite{Centrella84}], an enhancement of the
 extrinsic curvature  of order $\rho L^2/\tau$ will
occur.
      From the Hamiltonian constraint [Eq. (A3)  of \cite{Centrella84}] the trace of the
  extrinsic curvature  will be reduced by a factor
the order of $(\delta K/K)^2$  and since $\dot a/a \sim K/3$, $\dot a/a$
becomes more nearly constant implying acceleration.

 \acknowledgments

The authors acknowledge useful discussions with H. Duan C. Kolda, N. Q. Lan and  M. Patel 
regarding possible candidates for decaying dark matter.
Work at Lawrence Livermore National Laboratory performed under the auspices of the U.S. Department of Energy under under contract
W-7405-ENG-48 and NSF grant PHY-9401636.  Work at the University of Notre Dame supported
by the U.S. Department of Energy under 
Nuclear Theory Grant DE-FG02-95-ER40934. Work at UCSD supported in part by NSF grant PHY-04-00359.

\end{document}